\documentclass[a4paper]{article}

\usepackage{RR}
\RRNo{6219}
\usepackage{algorithmic}
\usepackage{algorithm}
\usepackage[T1]{fontenc}
\usepackage[latin1]{inputenc}

\usepackage{amsmath}
\usepackage{amssymb}
\usepackage{amsthm}
\usepackage{nicefrac}

\theoremstyle{plain}
\newtheorem{theorem}{\sc{Theorem}}[section]
\newtheorem*{theoremst}{\sc{Theorem}}
\newtheorem{proposition}[theorem]{\sc{Proposition}}

\newtheorem{corollary}[theorem]{\sc{Corollary}}

\theoremstyle{definition}
\newtheorem{definition}[theorem]{Definition}

\newtheorem*{examples}{Examples}

\newtheorem{lemma}[theorem]{\sc{Lemma}}

\theoremstyle{remark}
\newtheorem{remarks}[theorem]{Remarks}

\newtheorem{remark}[theorem]{Remark}

\newtheorem*{notations}{Notations}

\newcommand{\eps}{\varepsilon}                        
\renewcommand{\phi}{\varphi}                            

\renewcommand{\bar}[1]{\overline{#1}}
\renewcommand\over[2]{{\,\buildrel #1\over#2\,}}
\renewcommand{\leq}{\leqslant}
\renewcommand{\geq}{\geqslant}

\newcommand{\ie}{{\it{ie. }}}
\newcommand{\cf}{{\it{cf. }}}

\newcommand{\mc}[1]{\ensuremath{\mathcal{#1}}}

\newcommand{\nr}[1]{\left\Vert #1\right\Vert}         
\newcommand{\abs}[1]{\left\vert #1\right\vert}        
\DeclareMathOperator{\interieur}{int}             
\DeclareMathOperator{\diam}{diam}

\DeclareMathOperator{\Lip}{Lip}
\newcommand{\SLip}{\mathrm{Lip}}
\DeclareMathOperator{\Vor}{Vor}

\newcommand{\D}{\mathcal{D}}                         
\DeclareMathOperator{\BV}{BV}
\DeclareMathOperator{\var}{V}

\newcommand{\tq}{\,;\,}                          
\newcommand{\restr}[2]{\left.#1\right|_{#2}}         

\newcommand{\sur}[3][.5ex]{%
  \raise#1\hbox{\ensuremath{#2}}\!\!\left/%
    \vphantom{\raise#1\hbox{\ensuremath{#2}}}\!%
    \lower#1\hbox{\ensuremath{#3}} \right.%
}

\DeclareMathOperator{\proj}{proj}                    
                    % ...

\newcommand{\M}{\mathcal{M}}                         
\newcommand{\orth}{\bot}                             
\newcommand{\sca}[2]{\langle #1 | #2\rangle}         

\DeclareMathOperator{\supp}{supp}                    

\newcommand{\T}{\mathrm{T}}          
\renewcommand{\d}{\mathrm{d}}          

\newcommand{\R}{\mathbb{R}}                          

\newenvironment{narrow}[2]{%
 \begin{list}{}{%
  \setlength{\topsep}{0pt}%
  \setlength{\leftmargin}{#1}%
  \setlength{\rightmargin}{#2}%
  \setlength{\listparindent}{\parindent}%
  \setlength{\itemindent}{\parindent}%
  \setlength{\parsep}{\parskip}%
 }%
\item[]}{\end{list}}

\newsavebox{\fmbox}
\newenvironment{cadreth}
     {\begin{lrbox}{\fmbox}
            \begin{minipage}{0.9\textwidth}}
     {      \end{minipage}
      \end{lrbox}
      \fbox{\usebox{\fmbox}}} 

\newenvironment{cadrefg}
     {\begin{lrbox}{\fmbox}
            \begin{minipage}{\textwidth}}
     {      \end{minipage}
      \end{lrbox}
      \fbox{\usebox{\fmbox}}}

\RRdate{14 June 2007}
\RRauthor{Frédéric Chazal \and David Cohen-Steiner \and Quentin Mérigot}
\authorhead{Chazal, Cohen-Steiner \& Mérigot}

\RRtitle{Stabilité de mesures de bord}
\RRetitle{Stability of boundary measures}

\titlehead{Boundary measures}

\RRresume{
Nous introduisons la notion de \emph{mesure de bord} d'échelle $r$
d'un sous-ensemble compact de l'espace euclidien de dimension
$n$. Nous montrons comment calculer ces mesures pour un nuage de
points et suggérons que ces mesures peuvent être utilisées pour de la
détection de \emph{features}. La principale contribution de ce travail
est la démonstration d'un théorème quantitatif de stabilité des
mesures de bord, utilisant des outils de l'analyse convexe et de 
la théorie géométrique de la mesure. En corollaire, wous obtenons un
résultat de stabilité des \emph{mesures de courbure} d'un
compact (notion introduite par Federer), permettant de les calculer à
partir d'approximations du compact par des nuages de points. 
}
\RRabstract{
We introduce the \emph{boundary measure} at scale $r$ of a compact subset
of the $n$-dimensional Euclidean space. We show how it can
be computed for point clouds and suggest these measures can be used
for \emph{feature detection}. The main contribution of this work is
the proof a quantitative stability theorem for boundary measures
using tools of convex analysis and geometric measure theory. As a
corollary we obtain a  stability result for Federer's
\emph{curvature measures} of a compact, allowing to compute them from
point-cloud approximations of the compact.
}
\RRmotcle{détection de dimension, nuages de points, mesures de
  courbure, fonctions convexes, plus proche voisin.}
\RRkeyword{dimension detection, point clouds, curvature measures,
  convex functions, nearest neighbor.}

\RRprojet{Geometrica}
\RRtheme{\THSym}
\URSophia

\usepackage{graphicx}

\DeclareMathOperator{\reach}{reach}
\DeclareMathOperator{\law}{law}
\DeclareMathOperator{\mass}{mass}

\renewcommand{\L}{\mathrm{L}}
\DeclareMathOperator{\ddiv}{div}
\newcommand{\E}{\mathbb{E}}
\newcommand{\W}{\mathcal{W}}
\newcommand{\Ck}{\mathcal{C}}
\renewcommand{\H}{\mathcal{H}}

\newcommand{\Lloc}{\mathrm{L}^1_{\mathrm{loc}}}
\renewcommand{\ll}{\mathrm{length}}
\newcommand{\LL}{\mathcal{L}}
\newcommand{\NN}{\mathcal{N}}
\renewcommand{\D}{\mathcal{D}}
\newcommand{\p}{\mathrm{p}}
\newcommand{\dFM}{d_{\mathrm{FM}}}
\newcommand{\dbL}{d_{\mathrm{bL}}}

\begin{document}


\makeRR

\section*{Introduction}

\textbf{Motivations and previous work.}
The main goal of our work is to develop a framework for \emph{features
detection}: finding the boundaries, sharp edges, corners of a
compact set $K \subseteq \R^n$ knowing only a possibly noisy
point cloud sample of it.

This problem has been an area of active research in computer science
for some years. Many of the currently used methods for feature and
dimension detection (see \cite{dey2003sda} and the references therein)
rely on the computation of a Voronoï diagram. The cost of this computation
is exponential in the dimension and cannot be practically realized for
an ambient dimension much greater than three.
In low dimension, several methods have been invented for boundary
detection (mostly to detect holes), for example \cite{funke2006hdo}
(2D, graph-based), \cite{bendels:dhp} (3D), and
\cite{RosBroBroKimMLPR06}. Sharp edges detection has also been studied
in \cite{gumhold2001fep}, and recently in \cite{dhos07}. 

The algorithms we develop have three main advantages: they are built
on a strong mathematical theory, are robust to noise and their cost
depend only on the intrinsic dimension of the sampled compact
set. None of the existing methods for feature detection share these
three desirable properties at the same time.


 
\medskip
\textbf{Boundary measures and their stability.}
Given a scale parameter $r$, we associate to each compact subset $K$
of $\R^n$ a probability measure $\beta_{K,r}$.
This \emph{boundary measure} of $K$ \emph{at scale $r$} as we call it,
gives for every Borel set $A \subseteq \R^n$ the probability that the
projection on $K$ of a random point at distance at most $r$ of $K$
lies in $A$ (the \emph{projection on $K$}, denoted by $\p_K$, maps
almost any point in $\R^n$ to its closest point in $K$).  

Intuitively, the measure $\beta_{K,r}$ will be more
concentrated on the \emph{features} of $K$: for instance, if $K$ is a
convex polyhedron in $\R^3$, $\beta_{K,r}$
will charge  the edges more than the faces, and the vertices even more
(see example \ref{bm:examples}). It should also be noticed that this 
measure is closely related to Federer's \emph{curvature measures}
(introduced in \cite{federer1959cm}). 

This article focuses on the stability properties of the boundary
measures, showing that they can be approximated from a noisy
sample of $K$. The problem of extracting geometric information from
these boundary measures will be treated in an upcoming work. The main
stability theorem can be stated as follow:
\begin{theoremst}[\ref{th:stab}]
If one endows the set of compact subsets of $\R^n$ with the Hausdorff
distance, and the set of compactly supported probability measures on
$\R^n$ with the Wasserstein distance, the map 
$ K \mapsto \beta_{K,r}$ is locally $1/2$-Hölder.
\end{theoremst}
In the sequel we will make this statement more precise by giving
explicit constants. A very similar stability result for a
generalization of Federer's \emph{curvature measures} is deduced from
this theorem. We deduce theorem \ref{th:stab} from the two theorems
\ref{th:grad} and \ref{th:off} below, which are also interesting in
their own. 

\begin{theoremst}[\ref{th:grad}]
Let $E$ be an open subset of $\R^n$ with $(n-1)$--rectifiable
boundary, and $f,g$ be two convex functions such that $\diam(\nabla f(E) \cup
\nabla g(E)) \leq k$. Then there exists a constant $C(n,E,k)$ depending
only on $n$ and $E$ such that for $\nr{f-g}_\infty$ small enough,
 $$ \nr{\nabla f - \nabla g}_{\mathrm{L}^1(E)} \leq
 C(n,E,k)\nr{f-g}_\infty^{1/2} $$ 
\end{theoremst}

\begin{theoremst}[\ref{th:off}]
If $K$ is a compact set of $\R^n$, for every positive $r$,
$\partial K^r = \{ x \tq \d(x,K) = r\}$ is
$(n-1)$--rectifiable and 
$ \mc{H}^{n-1}(\partial K^r) \leq \NN(\partial K, r)\times \omega_{n-1}(2 r)$
\end{theoremst}

Theorem \ref{th:grad} is used to show that the map $K \mapsto
\p_K \in \L^1(E)$ (where $\p_K$ is the projection on $K$) is locally
$1/2$-Hölder, which is the main ingredient for the stability result. 
Theorem \ref{th:off} improves upon \cite{oleksiv1985fhm}, in which
Oleksiv and Pesin  prove the finiteness of the measure of the
level sets of the distance function to $K$. It is used here as a tool
to show that $K^r \Delta {K'}^r$ is small when $K$ and $K'$ are close
($A \Delta B$ being the symmetric difference between $A$ and $B$, and
$K^r$ being the set of points at distance at most $r$ from $K$).

\medskip
\textbf{Outline.} In the first section we give some examples of
boundary measures and show how they can be computed efficiently for
point clouds. The second and third sections contain the proofs of
theorems \ref{th:off} and \ref{th:grad} respectively. In the
fourth section we deduce from these theorems the stability results for
boundary and curvature measures.

\section{Definition of boundary measures}

\subsubsection*{Some examples of boundary measures}
\begin{notations}
If $K$ is a compact subset of $\R^n$, the distance to $K$ is defined
as $\d_K(x) = \min_{y \in K} \nr{x-y}$. The $r$-tubular neighborhood
or $r$-offset around a subset $F \subseteq \R^n$ is the set of points
at distance at most $r$ from $F$, and is denoted by $F^r$.

For $x \in \R^n$, the set of points $y \in K$ that realizes this
minimum is denoted by $\proj_K(x)$. One can show that $\#
\proj_K(x) = 1$ iff $\d_K$ is differentiable at $x$. Since $\d_K$ is
$1$-Lipschitz, a theorem of Rademacher ensures that both conditions
are true for almost every point $x\in\R^n$.  

This allows us to define a function $p_K \in \Lloc(\R^n)$, called the
projection on $K$, which maps (almost) every point $x \in \R^n$ to its
only closest point in $K$. The $s$-dimensional Hausdorff measure is
denoted by $\H^s$ ; in particular $\H^n$ coincides with the usual
Lebesgue measure on $\R^n$.
\end{notations}

\begin{definition}
The $r$-scale boundary measure $\beta_{K,r}$ of a compact $K$ of
$\R^n$ associates to any Borel set $A \subseteq \R^n$ the probability
that the projection of a random point at distance less than $r$ of $K$ lies
in $A$. 

If we denote by $\mu_{K,r}$ the pushforward of the uniform measure
on $K^r$ by the projection on $K$, \ie for all Borel set $A \subseteq
\R^n$, $\mu_{K,r}(A) = \H^n(p_K^{-1}(A) \cap K^r)$, then $\beta_{K,r}
= \H^n(K^r)^{-1} \mu_{K,r}$.
\end{definition}

\begin{examples}
\label{bm:examples}
\begin{enumerate}
\item  If $C = \{x_i; 1\leq i \leq N\}$ is a «point cloud», that is a finite
set of points of $\R^n$, then $\beta_{C,r}$ is a sum of weighted Dirac
measures. Indeed, if $\Vor_C(x_i)$ denotes the Voronoi cell of
$x_i$, that is the set of points closer to $x_i$ than to any
other point of $C$, we have
$$\mu_{C,r} = \sum_{i=1}^n \H^n(\Vor_C(x_i) \cap C^r) \delta_{x_i}$$
\item Let $S$ be a unit-length segment in the plane with endpoints $a$
  and $b$. The set $S^r$ is the union of a rectangle of dimension $1
  \times 2 r$ whose points projects on the segment and
  two half-disks of radius $r$ whose points are projected on $a$ and $b$. It
  follows that 
$$\mu_{S,r} = 2r \restr{\H^1}{S} + \frac{\pi}{2}
  r^2 \delta_a + \frac{\pi}{2} r^2 \delta_b$$
\item Let $P$ be a convex solid polyhedron of $\R^3$,  $\{ e_j \}$ be
  its edges and $\{ v_k \}$ be its vertices. We  
  denote by $a(e_j)$ the angle between the normals of the two faces containing
  $e_i$, and by $K(v_k)$ the solid angle formed by the normal cone at 
  $v_k$. Then one can see that  
$$
\mu_{P,r} = \restr{\H^3}{P}  + r \restr{H^2}{\partial P} + \sum_j r^2 a(e_j) \times \restr{\H^1}{e_j} 
+ \sum_k r^3 K(v_k) \delta_{v_k}
$$

\item More generally, if $K$ is a compact with positive reach, in the
  sense that there exists a positive $r$ such that the projection on
  $K$ is unique for any point in $K^r$, there exist Borel measures
  $(\Phi_{K,i})_{0 \leq  i \leq n}$ on $\R^n$ such that 
$$ \mu_{K,r} = \sum_{i=0}^n r^{n-i} \omega_{n-i}
\Phi_{K,i}$$
where $\omega_i$ is the volume of the unit sphere in $\R^{i+1}$. These
measures $\Phi_{K,i}$ are called the \emph{curvature measures} of the
compact set $K$ and have been introduced under this form by Federer in
\cite{federer1959cm}, generalizing existing notions in the case of
convex subsets and compact smooth submanifolds of $\R^n$ (Minkowski's
\emph{Quermassintegral} and Weyl's tube formula, \cf \cite{weyl1939vt}).
\end{enumerate}
\end{examples}

\begin{figure}[ht]
\centering\includegraphics[height=5cm]{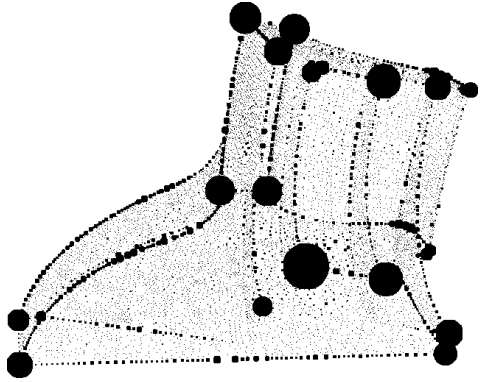}
\caption{Boundary measure for a sampled mechanical part.}
\label{fig:corner}
\end{figure}

The second and third  examples show exactly the kind of behaviour we
want to exhibit (and so does figure \ref{fig:corner}): the measure
$\beta_{K,r}$ can be written as a sum of weighted Hausdorff measures
of various dimension, concentrated on the features of $K$: its
boundary, its edges and its corners. This remark together with the
stability theorem for boundary measures shows that they are a suitable
tool to be used in robust feature extraction algorithms. In the next
paragraph we show how to compute them efficiently for point clouds.



\subsubsection*{The boundary measure of a point cloud}

A fast method for computing the boundary measures of point clouds is
of crucial importance for practical applications. Indeed, most
real-world data, either $3$D (laser scans) or higher dimensional is
given in the form of an  unstructured point cloud. Since computing
the Voronoï diagram of a point cloud has an exponential cost in the 
\emph{ambient} dimension, we will be using a probabilistic Monte-Carlo
method to get an approximation of the boundary measures. In a very
general way, if $\mu$ is an absolutely continuous measure on $\R^n$,
one can compute $\p_{\#C} \mu$ as shown below. The three main steps of
this algorithm (\textbf{I}, \textbf{II}, and \textbf{III}) are
described with more detail in the following paragraphs.  

\vspace{.5cm}
\begin{algorithmic}
\STATE\textbf{Input:} a point cloud $C=\{x_i\}$, a measure $\mu$
\STATE\textbf{Output:} an approximation of $\p_{C\#} \mu$ in the form
$\sum k(i) \delta_{X_i}$ 
\STATE[\textbf{I.}] Choose $N$ big enough to get a good approximation
with high confidence 
\WHILE {$n \leq N$}
   \STATE [\textbf{II.}] Choose a random point $X_n$ with probability
   distribution $\mu$ 
   \STATE [\textbf{III.}] Finds its closest point $x_i$ in the cloud $C$,
   add $1$ to $n(x_i)$ 
\ENDWHILE

\STATE \textbf{return} $\left[n(x_i)/N\right]_i$.
\end{algorithmic}
\vspace{.5cm}

\textbf{Step I.} The measure $\mu_N = 1/N \sum_{i\leq N} \delta_{X_i}$
where $(X_i)$ is a sequence of independent random variables whose law
are $\mu$ is called an \emph{empirical measure}. The question of
whether (and at what speed) $\mu_N$ converge to $\mu$ as $N$ grows to
infinity is well-known to probabilists and statisticians. The results
of this section are not original and can probably be improved,
they are presented here only to give \emph{proof-of-concept} bounds
for $N$.

\begin{theorem}[Hoeffding's inequality]
If $(Y_i)$ is a sequence of independent $[0,1]$-valued random variables
whose common law $\nu$ has a mean $m \in\R$, and $\bar{Y}_N = (1/N)
\sum_{i\leq N} Y_i$ then $$ \mathbb{P}(\abs{\bar{Y}_N - m} \geq \eps)
\leq 2 \exp(-2 N\eps^2)$$ 
\end{theorem}

In particular, let's consider a family $(X_i)$ of independent random
variables distributed according to the law $\p_{C\#} \mu$. Then, for
any $1$-Lipschitz function $f: \R^n \to \R$ with $\nr{f}_\infty
\leq 1$, one can apply Hoeffding's inequality to the family of random
variables $Y_i = f(X_i)$ : 
$$ \mathbb{P}\left[\abs{\frac{1}{N} \sum_{i=1}^N f(X_i) - \int f
    \d\mu} \geq \eps\right] \leq 2 \exp(-2 N \eps^2)$$ 
This kind of estimate also follows from Talagrand
$\T_1(\lambda)$-inequalities, in which case the factor $2$ in the
exponential is replaced by $2\lambda$. Bolley, Guillin and Villani use
this fact to get quantitative concentration inequalities for empirical
measures with non-compact support in \cite{bolley2007qci}.

We now let $\mathrm{BL}^1(C)$ be set of Lipschitz functions $f$ on
$C$ whose Lipschitz constant $\Lip f$ is at most $1$ and
$\nr{f}_\infty \leq 1$. We let
$\NN(\mathrm{BL}^1(C),\nr{.}_\infty,\eps)$ be the minimum
number of balls of radius at most $r$ (with respect to the
$\nr{.}_\infty$ norm)  needed to cover $\mathrm{BL}^1(C)$. Proposition
\ref{prop:Nlip} gives a bound for 
this number.  It follows from the definition of the bounded-Lipschitz
distance (see \ref{def:bl}) and from the union bound that 
$$\mathbb{P}\left[\dbL\left(\p_{C\#} \mu_N, \p_{C\#} \mu\right) \geq
  \eps \right]
\leq 2 \NN(\mathrm{BL}^1(C), \nr{.}_\infty, \eps/4) \exp(-N\eps^2/2)$$

\begin{proposition}
\label{prop:Nlip}
For any compact metric space $K$,
$$ \NN(\mathrm{BL}^1(K),\nr{.}_\infty,\eps) \leq
\left(\frac{4}{\eps}\right)^{\NN(K,\eps/4)}$$ 
\end{proposition}
\begin{proof}
Let $X=\{x_i\}$ be an $\eps/4$-dense family of points of $K$ with $\#X =
\NN(K,\eps/4)$. It is easily seen that for every $1$-Lipschitz
functions $f,g$ on $K$, $\nr{f-g}_\infty \leq
\nr{\restr{(f-g)}{X}}_\infty + \eps/2$. Then, one concludes using that 
$ \NN(\mathrm{BL}^1(X),\nr{.}_\infty,\eps/2) \leq
(4/\eps)^{\# X}$.
\end{proof}

\emph{In fine} one gets the following estimate on the
bounded-Lipschitz distance between the empirical and the real measure:
$$\mathbb{P}\left[\dbL\left(\p_{C\#} \mu_N, \p_{C\#} \mu\right) \geq
  \eps \right] \leq 2 \exp\left(\ln(16/\eps)\NN(C,\eps/16) - N
  \eps^2/2\right)$$ 

Since $C$ is a point cloud, the coarsest possible bound on
$\NN(C,\eps/16)$, namely $\#C$, shows that computing an
$\eps$-approximation of the measure $\p_\# \mu$ with high confidence
(\emph{eg.} $99\%$) can be done with $N = O(\#C\ln(1/\eps)/\eps^2)$. 
\smallskip

\textbf{Step II.}
To simulate the uniform measure on $K^r$ one cannot simply shoot
points in a bounding box of $K^r$, keeping those that are actually in
$K^r$ since this has an exponential cost in the ambient dimension.
Luckily there is a simple algorithm to generate points according to
this law which relies on picking a random point $x_i$ in the cloud $C$
and then a point $X$ in $B(x_i,r)$ --- taking into account the
overlap of the balls $B(x,r)$ where $x \in C$: 

\vspace{.5cm}
\begin{algorithmic}
\STATE\textbf{Input:} a point cloud $C=\{x_i\}$, a scalar $r$
\STATE\textbf{Output:} a random point in $C^r$ whose law is $\restr{\H^n}{K^r}$

\REPEAT
   \STATE Pick a random point $x_i$ in the point cloud $C$
   \STATE Pick a random point $X$ in the ball $B(x_i,r)$
   \STATE Count the number $k$ of points $x_j \in C$ at distance at most $r$
   from $X$
   \STATE Pick a random integer $d$ between $1$ and $k$
\UNTIL $d = 1$

\STATE \textbf{return} $X$.
\end{algorithmic}
\vspace{.5cm}

\textbf{Step III.} 
The trivial algorithm for computing the projection of a point on a
point cloud takes exactly $n$ steps. Since generally $N$ will an order
of magnitude greater than $n$ we might improve the overall $O(n^2)$
cost by maintaining a data structure which allows fast
nearest-neighbour queries. This problem is notoriously difficult and
until recently most of the efficient algorithms in high dimension were
only able to compute \emph{approximate nearest neighbours}. This
amounts to replacing $\p_C$ by a map $\tilde{\p}_\eps$ with the
property that for all $x$, $\nr{\tilde{\p}_\eps(x) - \p_C(x)} \leq
(1+\eps) \d_C(x)$. Unfortunately, the techniques we develop in this
paper do not seem to apply directly to get quantitative closeness
estimates for 
the measures $\tilde{\p}_{\eps\#}\mu$ and $\p_{K\#}\mu$. 

It should be noted that for low entropy point clouds, nearest neighbor
queries can be done more efficiently. For instance, a recent article
by Beygelzimer, Kakade and Langford (\cf  \cite{beygelzimer2006ctn})
introduces a structure called \emph{cover   trees} which allows an
\emph{exact} nearest neighbour query with 
complexity $O(c^{12} \log n)$ where $c$ is related to the intrinsic
dimension of the point cloud, with an initialisation cost of $O(c^6 n
\log n)$.

\subsubsection*{Wasserstein distance and stability}

Since our goal is to give a quantitative stability result for boundary
measures, we need to put a metric on the space of probability measures
on $\R^n$. The Wasserstein distance, related to the Monge-Kantorovich
optimal transportation problem seemed intuitively (and later happened
to really be) appropriate for our purposes. A good reference on
this topic is Cédric Villani's book \cite{villani2003tot}.

\begin{definition}
\label{def:bl}
The set of measures (resp. probability measures) on $\R^n$ is denoted by
$\M(\R^n)$ (resp. $\mc{M}^1(\R^n)$). We endow $\M(\R^n)$ with the
bounded Lipschitz distance, \ie
$$\forall \mu,\nu \in \M(\R^n), \dbL(\mu, \nu) =
\sup_{\nr{\phi}_\SLip\leq 1} \abs{\int \phi \d\mu - \int \phi \d\nu} $$ 
where the supremum is taken over all Lipschitz functions $\phi$ with 
$\nr{\phi}_\SLip = \Lip \phi + \nr{\phi}_\infty \leq 1$ ($\Lip \phi$ being
the smallest constant $k$ such that $\phi$ is $k$-Lipschitz).

We put two distances on $\M^1(\R^n)$ (which are in fact identic, see
below). The Fortet-Mourier distance, which is almost the same as the
bounded Lipschitz one: 
$$\forall \mu,\nu \in \M^1(\R^n), \dFM(\mu, \nu) =
\sup_{\Lip \phi \leq 1} \abs{\int \phi \d\mu - \int \phi \d\nu} $$ 

And the  Wasserstein distance:  
$$ \W_1(\mu,\nu) = \inf \left\{ \E(\d(X,Y)) \tq \law(X) = \mu,~ \law(Y)
  = \nu \right\} $$
where the infimum is taken over all random variables $X$ and $Y$ whose
laws are $\mu$ and $\nu$ respectively.
\end{definition}

\begin{notations}
If $\mu$ and $\nu \in \M(\R^n)$ are absolutely continuous with respect
to $\H^n$, \ie $\d\mu = \phi \d \H^n$ and $\d\nu = \psi \d \H^n$ we
denote by $\mu \cap \nu$ the measure defined by $\d(\mu\cap\nu) =
\min(\phi,\psi) \d \H^n$, and $\mu \Delta \nu = \mu + \nu - 2 \mu \cap
\nu$. 
\end{notations}

\begin{proposition}
If $\mu \in \M(\R^n)$ is absolutely continuous with respect to the
Lebesgue measure, and $f,g: \R^n \to\R^n$ are two functions in
$\L^1(\mu)$, then $$ \dbL(f_\# \mu, g_\# \mu) \leq \nr{f -
  g}_{\L^1(\mu)}$$ 

If $\mu$ and $\nu$ are two absolutely continuous measures on $\R^n$,
$$ \dbL(f_\# \mu, g_\# \nu) \leq \nr{f - g}_{\L^1(\mu\cap\nu)} +
\mass(\mu\Delta\nu) $$
\end{proposition}

\begin{proof}
For any $1$-Lipschitz function $\phi$ on $\R^n$, 
$$
\begin{aligned}
\abs{\int \phi \d f_\# \mu - \int \phi \d g_\# \mu} &=
\abs{\int \phi\circ f \d\mu
- \int \phi\circ g \d\mu}\\
&\leq \Lip\phi \int \nr{f-g} \d\mu \leq \nr{f-g}_{\L^1(\mu)}
\end{aligned}$$

For the second inequality, let us first remark that there exists two
positive measures $\mu_r$ and $\nu_r$ such that $\mu = \mu \cap \nu +
\mu_r$ and $\nu = \mu \cap \nu + \nu_r$. Then,
$$ \dbL(f_\# \mu, g_\# \nu) \leq \dbL(f_\# \mu, f_\# \mu \cap \nu) +
\dbL(f_\# \mu \cap \nu, g_\# \mu \cap \nu) + \dbL(g_\# \mu, g_\# \mu
\cap \nu) $$

Now let us bound one of the extreme terms of the sum,
$$\forall \phi \hbox{ s.t} \nr{\phi}_\infty \leq 1,
\abs{\int \phi \d f_\# \mu - \int \phi \d f_\# \mu \cap \nu} = \abs{\int \phi
\circ f \d \mu_r} \leq  \mass(\mu_r) $$
One concludes using that $\mu_r + \nu_r = \mu \Delta \nu$.
\end{proof}

\begin{corollary}
If $K$ and $K'$ are two compact subsets of $\R^n$,
$$ \dbL(\mu_{K,r}, \mu_{K',r}) \leq \nr{p_K - p_K'}_{\L^1(K^r \cap
  {K'}^r)} + \H^n(K^r \Delta {K'}^r)$$
\label{coro:triangle}
\end{corollary}

Hence to get a quantitative continuity estimate for the map $K \mapsto
\mu_{K,r}$ one needs to show that if $K$ and $K'$ are Hausdorff-close, 
$K^r \Delta {K'}^r$ is small, and to evaluate the continuity modulus
of $K \mapsto \p_K \in \L^1(K^r\cap {K'}^r)$. This is the purpose of the two
following paragraphs.  

\section{$K^r \Delta {K'}^r$ is small when $K$ and $K'$ are close}

It is not hard to see that if $\d_H(K, K')$ is smaller than $\eps$,
then $K^r \Delta {K'}^r$ is contained in $(K^{r+\eps} \setminus
K^{r-\eps})$. The volume  of this thick tube around $K$ can then be
expressed as an integral of the area of the hypersurfaces  $\partial
K^t$. 

The next proposition gives a bound for the measure of the $r$-level
set $\partial K^r$ of a compact set $K\subseteq \R^n$ depending only
on its covering number $\NN(K,r)$ (\ie the minimal number of closed balls
of radius $r$ needed to cover $K$).  In what follows,
$K^{r}$ is the set of points of $\R^n$ at distance less than $r$
of $K$, and $\partial K^r$ is the boundary of this set, \ie the
$r$-level set of $\d_K$. In this paragraph, we prove the following
theorem : 

\begin{theoremst}
If $K$ is a compact set of $\R^n$, for every positive $r$,
$\partial K^r$ is $\mc{H}^{n-1}$--rectifiable and
$\mc{H}^{n-1}(\partial K^r) \leq \NN(\partial K,r)\times
\omega_{n-1}(2r)$ 
\end{theoremst}

This proposition improves over a result of finiteness of the level sets of
the distance function to a compact set, proved by by Oleksiv and
Pesin in \cite{oleksiv1985fhm}. We begin by proving it in the
special case of ``$r$-flowers''. A $r$-flower $F$ is the the boundary of
the $r$-tube of a compact set contained in a ball $B(x,r)$, \ie
$F = \partial K^r$ where $K \subseteq B(x,r)$. The difference
with the general case is that if $K \subseteq B(x,r)$, then $K^r$ is a
star-shaped set with respect to $x$. Thus we can define a ray-shooting
application $s_K: \mc{S}^{n-1} \to \partial K^r$ which maps any $v \in
\mc{S}^{n-1}$ to the intersection of the ray emanating from $x$ with
direction $v$ with $\partial K^r$.  

\begin{figure}[ht]
\centering\includegraphics[height=4cm]{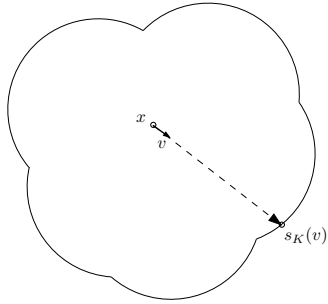}
\caption{Ray-shooting from the center of a flower.}
\end{figure}

\begin{lemma}
Let $K = \{ e \} \subseteq B(x,r)$ and define $s_e$ as above. Then $s_e$
is $2 r$-Lipschitz (with respect to the sphere's inner metric) and its
Jacobian is at most $(2 r)^{n-1}$.  
\end{lemma}

\begin{proof}
Solving the equation $\nr{x+tv - e} = r$ with $t \geq 0$ gives 
$$ s_e(v) = x + \left(\sqrt{\sca{v}{x-e}^2 +
 r^2 - \nr{x-e}^2} - \sca{v}{x-e}\right)v$$

Denote by $H_v$ the orthogonal of the $2$-plane $P$ spanned by $v$ and
$s_e(v)-e$. For each vector $w$ chosen in $H_v$, a simple calculation
gives: 
$$ s_e(v+tw) = s_e(v) + t w \nr{s_e(v) - x} + o(t^2)$$
Hence the derivative of $s_e$ along $H_v$ is simply the multiplication
by $\nr{s_e(v) - x} \leq 2 r$.  

Now, we now consider the case of the $2$-plane $P$. We denote by
$\theta$ the angle between $s_e(v)-x$ and $s_e(v)-e$ and by $w$ a vector
tangent to $v$ in the intersection of the sphere with $P$. Then
$$\frac{\nr{(\d s_e)_v(w)}}{\nr{w}} =
\frac{\nr{s_e(v)-x}}{\abs{\cos(\theta)}}$$ 
Now let us remark that 
$$\begin{aligned}\nr{s_e(v)-e}\nr{s_e(v)-x} \abs{\cos(\theta)} &=
\abs{\sca{s_e(v)-e}{s_e(v)-x}} \\
&= \frac{1}{2}(\nr{x -
  s_e(v)}^2 + \nr{s_e(v)-e}^2 - \nr{x-e}^2) \\
& \geq \frac{1}{2}\nr{x - s_e(v)}^2
\end{aligned}$$ 
Finally we have proved that $ \nr{(\d s_e)_v} \leq 2 r$. The
result follows by integration.
\end{proof}

We denote by $\omega_{n}(r)$ the $n$-Hausdorff measure of the
$n$-sphere of radius $r$. 

\begin{corollary}
A $r$-flower in $\R^n$ is a $\mc{H}^{n-1}$--rectifiable set and its
measure is at most $\omega_{n-1}(2r)$.
\label{prop:flower}
\end{corollary}

\begin{proof}
Let $K \subseteq B(x,r)$ be the compact set generating the flower
$\partial K^r$. As above, for any vector $v \in \mc{S}^{n-1}$, we
denote by $s$ the  intersection of the ray $\{x + tv \tq t > 0\}$ with
$\partial K^r$. Since $K^r$ is a star-shaped set around $x$, $s$ is a
bijection from $\mc{S}^{n-1}$ to $\partial K^r$.

Now let $(y_k)$ be a dense sequence in $K$, and denote by $s_k$ the
projection from $\mc{S}^{n-1}$ to the flower $\partial (\cup_{i \leq
  k} \{y_i\})^r$  
defined as above. Then $(s_k)$ converges simply to $p$ on
$\mc{S}^{n-1}$. Indeed, if we fix $v \in \mc{S}^{n-1}$ and $\eps > 0$,
the segment joining $x$ and $s(v)$ truncated at a distance $\eps$ of
$s(v)$ is a compact set contained in $\interieur{K^r}$. It is covered
by the union $\cup_i B(y_i, r)$, so that for $N$ big enough it is
also covered by $\cup_{k \leq N} B(y_k,r)$. For those $N$, $\nr{s_k(x)
  - s(x)} \leq \eps$. 

Finally, $\partial K^r$ is the image of the sphere by $p$, which is $2
r$-Lipschitz as a simple limit of $2 r$-Lipschitz functions. 
\end{proof}

We now deduce a general bound on the measure of the tube boundary
$\partial K^r$ around a general compact set $K$ by covering it with a family of
flowers: 

\begin{theorem}
\label{th:off}
If $K$ is a compact set of $\R^n$, for every positive $r$,
$\partial K^r$ is a $\mc{H}^{n-1}$-rectifiable subset of $\R^n$ and
moreover, 
$$ \mc{H}^{n-1}(\partial K^r) \leq
\NN(\partial K,r)\times \omega_{n-1}(2 r)$$ \label{pr:vol:nh}
\end{theorem}
\begin{proof}
It is easy to see that $\partial K^r \subseteq \partial(\partial
K^r)$. Thus, if we let $(x_i)$ be an optimal covering of $\partial K$ by open
balls of radius $r$, and denote by $K_i$ the (compact) intersection of
$\partial K$ with $B(x_i,r)$, the boundary $\partial K^r$ is contained
in the union $\cup_i \partial K_i^r$. Hence its Hausdorff measure does
not exceed the sum $\sum_i \mc{H}^{n-1} (\partial K_i^r)$. One
concludes by applying the preceding lemma. 

\end{proof}

\begin{remark}
\begin{enumerate}
\item The bound in the theorem is tight, as one can check taking $K =
  B(0,r)$. 
\item
Let us notice that for some constant $C(n)$, $\NN(B(0,1), r)
\leq 1+C(n) r^{-n}$. From this and the above bound it follows that 
$$
\begin{aligned}
\mc{H}^{n-1}(\partial K^r) &\leq
(1+C(n)\times (\diam(K)/r)^{n}) \omega_{n-1}(2 r)\\
&\leq C'(n) \times (1+\frac{\diam(K)^n}{r})\end{aligned}$$
for some universal constant $C'(n)$ depending only on the ambient
dimension $n$. This last inequality was the one proved in
\cite{oleksiv1985fhm}.
\end{enumerate}
\end{remark}

To conclude we use a weak formulation of the \emph{co-area formula}, a
standard result of geometric measure theory (\cite{degiorgi1954tgd},
\cite{federer1959cm}), which reads
$$\int_{\R^n} \abs{\nabla_x f} \d\H^n(x) = \int_\R \H^{n-1}(f^{-1}(y))
\d \H^1(y) $$
whenever $f: \R^n\to \R$ is a Lipschitz map. From this formula and the
previous estimation follows that 

\begin{corollary}
\label{coro:szsym}
For any compact sets $K, K' \subseteq \R^n$, with $\d_H(K,K') \leq
\eps$, 
 $$
 \begin{aligned}
 \mc{H}^n(K^r \Delta {K'}^r) &
\leq \int_{r - 
   \eps}^{r+\eps} \mc{H}^{n-1}(\partial K^t) \d t\\
 &\leq 2 \NN(K,r-\eps) \omega_{n-1}(2r+2\eps) \times  \eps
 \end{aligned}$$


\end{corollary}

\section{The map $K \mapsto p_K$ is locally $1/2$-Hölder}

We now study the continuity modulus of the map $K \mapsto p_K \in
\L^1(E)$, where $E$ is a suitable open set.
We remind the reader of two well-known facts of convex analysis (see
for instance \cite{clarke1983oan}): 
\begin{enumerate}
\item If $f: \Omega \subseteq \R^n \to \R$ is a locally convex
  function, its subdifferential at a point $x$, denoted by $\partial_x
  f$ is the set of vectors $v$ of $\R^n$ such that for all $h \in
  \R^n$ small enough, $f(x+h) \geq f(x) +  
  \sca{h}{v}$. Then $f$ admits a derivative at $x$ iff
  $\partial_x f = \{ v \}$ is a singleton, in which case $\nabla_x f =
  v$.
\item A locally convex function has a derivative almost everywhere.
\end{enumerate}

\begin{lemma}
\label{lem:proj}
The function $v_K: \R^n \to \R, x \mapsto \nr{x}^2 - \d_K(x)^2$ is
convex with gradient $\nabla v_K = 2 p_K$ almost everywhere.
\end{lemma}

\begin{proof}
By definition, $v_K(x) = \sup_{y \in K} \nr{x}^2 - \nr{x-y}^2 = 
\sup_{y \in K} v_{K,y}(x)$ with $v_{K,y}(x) = 2\sca{x}{y} -
\nr{y}^2$. Hence $v_K$ is convex as a 
supremum of affine functions. Because $v_{K,p_K(x)}$ and $v_K$ take the same
value at $x$, $\partial_{x} v_{K,p_K(x)} = \{ 2 p_K(x) \} \subseteq
\partial v_K$. Since $v_K$ is differentiable almost everywhere,
equality must be true almost everywhere which concludes the proof.
\end{proof}

This lemma shows that $\nr{p_K - p_{K'}}_{\L^1(E)} = 1/2 \nr{\nabla
  v_K - \nabla v_{K'}}_{\L^1(E)}$. Our estimation of the continuity
modulus of the map $K \mapsto \p_K$ will follow from a general theorem
which asserts that if $\phi$ and $\psi$ are two uniformly close convex
functions with bounded gradients then $\nabla \phi$ and $\nabla
\psi$ are $\L^1$-close. The next proposition below is the
$1$-dimensional version of this result, from which we then deduce the
general theorem. 

\begin{proposition}
\label{prop:conv:dim}
If $I$ is an interval, and $\phi:I\to\R$ and $\psi:I\to\R$ are two convex
functions such that $\diam(\phi'(I) \cup \psi'(I)) \leq k$, then letting
$\delta = \nr{\phi-\psi}_{\L^\infty(I)}$,
$$\int_I \abs{\phi' - \psi'} \leq 6\pi (\ll(I)+k+\delta^{1/2})\delta^{1/2}
$$ 
\end{proposition}

\begin{lemma}
Let $f: I \to \R$ be a nondecreasing function with $\diam \phi(I) \leq
k$. Then, if $F$ is the completed graph of $f$, \ie the set of points 
$(x,y) \in I \times \R$ such that $\lim_{x^-} \phi \leq y \leq
\lim_{x^+} \phi$, then $\H^n(F^r) \leq 3\pi (\ll(I)+k+r) \times r$.
\end{lemma}

\begin{proof}
Let $\gamma:[0,1]\to F$ be a continuous parametrization of $F$,
increasing with respect to the lexicographic order on $\R^2$. Then,
for any increasing sequence $(t_i) \in [0,1]$ and $(x_i, y_i) = \gamma(t_i)$, 
$$ \sum_i \nr{\gamma(t_{i+1}) - \gamma(t_i)} \leq \sum_i x_{i+1} - x_i +
y_{i+1}-y_i \leq 
\ll(I) + k$$

Hence $\ll(F) \leq \ll(I)+k$. Thus we can choose a  $1$-Lipschitz 
parametrization of $F$, $\tilde{\gamma}: [0,\ll(I)+k] \to
F$. Then for any positive $r$, the set $X = \{ 
\tilde\gamma(i\times r) \tq 0 \leq i \leq N\}$ with $N$ the upper integer
part of $(\ll(I)+k)/r$, is such that any point of $F$ is at distance
at most $r$ of $X$. Hence $F^r$ is contained in $X^{2r}$, implying that 
$\H^n(F^r) \leq N \pi (3r/2)^2 \leq 3\pi(\ll(I)+k+r)r$.
\end{proof}

\begin{proof}[Proof of proposition \ref{prop:conv:dim}]
Let $I = [a,b]$ and $J = [c,c+k]$ be such that $\phi'(I) \cup \psi'(I)
\subseteq J$. Without loss of generality we will suppose that
$\psi'(a) = \phi'(a) = c$ and $\psi'(b) = \phi'(b) = c+k$. With this
assumption, the completed graphs $\Phi$ and $\Psi$  of $\phi'$ and
$\psi'$ defined as above are two rectifiable curves joining $(a,c)$
and $(b,c+k)$. We let $V$ be the set of points $(x,y) \in \R^2$ lying
between those graphs; the quantity we want to bound is $\int_I
\abs{\phi' - \psi'} = \mathcal{H}^2(V)$.  

Let $\delta = \nr{\phi-\psi}_{\L^\infty(I)}$. For any point $p =
(x,y)$ in $V$, and any $\delta' > \delta$, the closed disk $D =
\bar{B}(p, \sqrt{2\delta'/\pi})$ of volume $2\delta'$  centered at $p$
cannot be  contained in $V$. Indeed if it were, then the difference
$\kappa = \phi - \psi$  
would increase too much around $p$: since $\kappa'$ has  a constant
sign on this segment, $$\abs{\kappa(x + 2\delta'/\pi) -
  \kappa(x-2\delta'/\pi)} = \int_{x -  2\delta'/\pi}^{x + 2\delta'/\pi}
\abs{\kappa'} \geq \H^2(D) = 2\delta' 
> 2 \delta$$
This contradicts $\nr{\kappa}_\infty = \delta$. Hence, $D$ must
intersects $\partial V$ implying that $V$ must be contained in $(\partial
V)^{\sqrt{2\delta'/\pi}}$ for any $\delta' > \delta$. Since $\partial
V = \Phi \cup \Psi$, the previous lemma gives 
$$\H^2(V) \leq \H^2\left(\Phi^{\sqrt{2\delta'/\pi}}\right) +
\H^2\left(\Psi^{\sqrt{2\delta'/\pi}}\right) \leq 6\pi(\ll(I) + k
+\sqrt{2\delta'/\pi}) \sqrt{2\delta'/\pi}$$ 
Letting $\delta'$ converge to $\delta$ concludes the proof.
\end{proof}

A generalization of this proposition in arbitrary dimension will
follow from an argument coming from integral geometry, \ie we will
integrate the inequality of proposition \ref{prop:conv:dim} over the
set of lines of $\R^n$ to get a bound on $\nr{\nabla \phi - \nabla
\psi}_{\L^1(E)}$.

We let $\LL^n$ be the set of oriented affine lines
in $\R^n$ seen as 
the submanifold of $\R^{2n}$ made of points $(u,p) \in \R^n \times
\R^n$ with $u \in  \mc{S}^{n-1}$ and $x$ in the hyperplane
$\{u\}^\orth$, and endowed with  the induced Riemannian metric. The
corresponding measure $\d\LL^n$ is invariant under rigid motions. We let
$\mathcal{D}_u^n$ be the set of oriented lines with a fixed direction
$u$. 

The usual Crofton formula (\cf \cite{morgan1988gmt} for instance)
states that for any $\H^{n-1}$--rectifiable subset $S$ of $\R^n$, with
$\beta_n$ the volume of the unit $n$-ball,
\begin{equation}
\label{eq:crof}
\H^{n-1}(S) = \frac{1}{2\beta_{n-1}} \int_{\ell
  \in \LL^n} \#(\ell \cap S) \d \ell
\end{equation}
where $\#X$ is the cardinality of $X$. We will also use the following
Crofton-like formula: if $K$ is a $\H^n$--rectifiable
subset of $\R^n$, 
\begin{equation}
\label{eq:crofbis}
\H^{n}(K) = \frac{1}{\omega_{n-1}} \int_{\ell \in \LL^n} \H^{1}(\ell \cap
K) \d \ell 
\end{equation}
which follows from the Fubini theorem (remember $\omega_{n-1}$ is the
volume of the $(n-1)$--sphere).

\begin{lemma}
Let $X: E \to \R^n$ be a $\L^1$-vector field on an open subset $E
\subseteq \R^n$. 
$$\int_E \nr{X} = \frac{n}{2\omega_{n-2}} \int_{\ell\in \LL^n}
\int_{y \in \ell\cap E} \abs{\sca{X(y)}{u(\ell)}} \d y \d \ell$$
\label{lem:intgeom}
\end{lemma}

\begin{proof}[Sketch of proof]
The family of vector fields of the form $\sum_i X_i
\chi_{\Omega_i}$, where the $\Omega_i$ are a finite number of disjoint
open subsets of $\R^n$ and $X_i$ are constant vectors, is $\L^1$-dense
in the space $\L^1(\R^n,\R^n)$. Using this fact and the continuity of
the two sides of the equality, it is enough to prove this
equality for $X = x \nr{X} \chi_E$ where $x$ is a constant unit
vector and $E$ a bounded open set of
$\R^n$. 

In that case, one has
$$\begin{aligned}
\int_{\ell \in \D_u^n} \int_{y \in \ell} \abs{\sca{X(y)}{u}} \d y \d
\ell &= \nr{X} \abs{\sca{x}{u}} \int_{\ell\in\D_u^n} \ll(E \cap \ell) \d
\ell\\ 
&
= \nr{X}_{\L^1(E)} \abs{\sca{x}{u}}
\end{aligned}$$
By a Fubini-like theorem one has
$$
\begin{aligned}
\int_{\ell \in \LL^n} \int_{y \in \ell} \abs{\sca{X(y)}{u(\ell)}} \d y \d
\ell &= \int_{u \in \mc{S}^{n-1}} \int_{\ell \in \D_u^n} \int_{y \in \ell}
\abs{\sca{X(y)}{u(\ell)}} \d y \d\ell \d u
\\ &=\nr{X}_{\L^1(E)} \int_{u \in \mc{S}^{n-1}} \abs{\sca{x}{u}} \d u
\end{aligned}$$
The last integral does, in fact, not depend on $x$ and its value can
be easily computed: 
$$
\begin{aligned}\int_{u \in \mc{S}^{n-1}} \abs{\sca{x}{u}} \d u 
&= 2 \omega_{n-2} \int_0^1 t (1-t^2)^{\frac{n}{2}-1}\d t\\
&= \frac{2}{n} \omega_{n-2}
\end{aligned}
$$
\end{proof}

\begin{theorem}
Let $E$ be an open subset of $\R^n$ with $(n-1)$--rectifiable
boundary, and $f,g$ be two locally convex functions on $E$ such that
$\diam(\nabla f(E) \cup \nabla g(E)) \leq k$. Then, letting $\delta =
\nr{f-g}_{\L^\infty(E)}$  $$ \nr{\nabla f - \nabla
  g}_{\mathrm{L}^1(E)} \leq C_1(n) (\H^n(E) +  (k+\delta^{1/2})
\H^{n-1}(\partial E)) \delta^{1/2} $$ with $C_1(n) \leq 6 \pi n$ as
soon as $n > 5$ (in fact, $C_1(n) = O(\sqrt{n})$). 
\label{th:grad}
\end{theorem}

\begin{proof}[Proof of the theorem]
The $1$-dimensional case follows from pro\-po\-sition
\ref{prop:conv:dim}: in that case, $E$ is a countable union of
intervals on which $f$ and $g$ satisfy exactly the hypothesis of the
proposition. Summing the inequalities gives the result with $C_1(1) = 6\pi$.

The general case will follow from this one with the use of integral
geometry. If we set $X = \nabla f - \nabla g$, $f_\ell =
\restr{f}{\ell \cap E}$ and $g_\ell = \restr{g}{\ell\cap E}$.
Lemma \ref{lem:intgeom} gives, letting $D(n) = n/(2\omega_{n-2})$,
$$
\begin{aligned}
\int_E \nr{\nabla f - \nabla g} &= D(n) \int_{\ell \in \LL^n}
\int_{y\in \ell\cap E} \abs{\sca{\nabla f - \nabla g}{u(\ell)}} \d y
\d\ell\\
&= D(n) \int_{\ell\in \LL^n} \int_{y\in\ell\cap E} \abs{f'_\ell - g'_\ell} \d y
\d\ell
\end{aligned}$$

The functions $f_\ell$ and $g_\ell$ satisfy the hypothesis of the
one-dimensional case, so that for each choice of $\ell$, and with 
$\delta = \nr{f-g}_{\L^\infty(E)}$,
$$\int_{y\in\ell\cap E} \abs{f'_\ell - g'_\ell} \d y
\leq  6 \pi D(n) (\H^1(E\cap \ell) + (k+\delta^{1/2}) \H^0(\partial
E\cap \ell)) \delta^{1/2}$$ 
It follows by integration on $\LL^n$ that
$$\int_E \nr{\nabla f - \nabla g} \leq 6\pi D(n) \left(\int_{\LL^n}
\H^1(E\cap \ell) \d\LL^n + (k+\delta^{1/2}) \int_{\LL^n}
\H^0(\partial E\cap \ell)\d\LL^n\right) \delta^{1/2}$$
The formula \ref{eq:crof} and \ref{eq:crofbis} show that the first
integral is equal (up to a constant) to the volume of $E$ and the
second to the $(n-1)$-measure of $\partial E$. This proves the theorem 
with $C_1(n) = 6 \pi D(n) (\omega_{n-1} + 2\beta_{n-1})$. To get the
bound on $C_1(n)$ one uses the formula $\omega_{n-1} = n \beta_{n}$ and
$\beta_{n+1} \leq \beta_n$ as soon as $n>5$.
\end{proof}

Multiplying $f$ and $g$ by the same positive factor $t$ and
optimizing the result in $t$ yields a better, homogeneous, bound : 

\begin{corollary}
\label{coro:homo}
Under the same hypothesis as in theorem \ref{th:grad}, one gets the
following bound, with $\delta = \nr{f-g}_{\L^\infty(E)}$ :
$$
\begin{aligned}
\nr{\nabla f - \nabla
  g}_{\mathrm{L}^1(E)} \leq 2 C_1(n) [(\H^n(E) \H^{n-1}(\partial E)&
\diam(\nabla f(E) \cup \nabla g(E)))^{1/2}\\
&+ \H^{n-1}(\partial
E)\delta^{1/2}] \delta^{1/2}
\end{aligned}$$
\end{corollary}

\begin{remark}
To get an homogeneous bound as in this corollary, one
could also optimize the one-dimensional bound of proposition
\ref{prop:conv:dim} before integrating on the set of affine lines of
$\R^n$ as in the proof of theorem \ref{th:grad}.
The bound obtained this way is always strictly better
than the ones of both theorem \ref{th:grad} and  corollary
\ref{coro:homo}, but involves an integral term
$$\int_{\ell \in \LL^n} \sqrt{\mathcal{H}^0(\ell\cap \partial E)
  \mathcal{H}^1(\ell\cap E)} \d\ell $$
whose intuitive meaning is not quite clear.
\end{remark}

Applying theorem \ref{th:grad} to the functions $v_K$ and $v_{K'}$ introduced
at the begining of this part and using lemma \ref{lem:proj}, one
easily gets : 
\begin{corollary}
\label{th:stab:E}
\label{coro:diffp}
If $E$ is an open set of $\R^n$ with rectifiable boundary, $K$ and
$K'$ two compact subsets of $\R^n$ then, with $R_K =
\nr{d_K}_{\L^\infty(E)}$ and $\eps = \d_H(K,K')$,
$$\begin{aligned}\nr{p_K - p_{K'}}_{\L^1(E)} \leq 
C_1(n) [\H^n(E)+ (\diam&(K)+ \eps + (2R_K + \eps)^{1/2} \eps^{1/2})
\H^{n-1}(\partial E)]\\ 
&\times (2R_K + \eps)^{1/2}\eps^{1/2}
\end{aligned}$$
In particular, if $\d_H(K,K')$ is smaller than $\min(R_K,
\diam(K), \diam(K)^2/R_K)$, there is another constant $C_2(n)$
depending only on $n$ such that
$$ \nr{p_K - p_{K'}}_{\L^1(E)} \leq C_2(n) [\H^n(E) + \diam(K)
\H^{n-1}(\partial E)] \sqrt{R_K\d_H(K,K')}$$
\end{corollary}


\begin{remarks}
\begin{enumerate}
\item This theorem gives in particular a quantitative version of the
 continuity theorem $4.13$ of $\cite{federer1959cm}$: if $(K_n)$ is a
 sequence of compact subsets of $\R^n$ with $\reach(K_n)\geq r > 0$,
 converging  to a compact set $K$, then $\reach(K) \geq r$ and
 $\p_{K_n}$ converges to $\p_K$ uniformly on each compact set
   contained in $\{ x \in \R^n \tq \d_K(x) < r \}$.
However we have to stress that the result we have proved is more
general since it does not make any assumption on the regularity of
$K_n$ --- at the expense of uniform convergence.
\item The second term of the bound involving $\H^{n-1}(\partial E)$  is
  necessary. Indeed, let us suppose that a bound  $\nr{p_K -
    p_{K'}}_{\L^1(E)} \leq C(K) \H^n(E)  \sqrt{\eps}$ were true
  around $K$ for any open set $E$. Now let $K$ be the union of two
  parallel hyperplane at distance $R$ intersected with a big sphere
  centered at a  point $x$ of their  medial hyperplane $M$. Let
  $E_\eps$ be a  ball of radius $\eps$ tangent to $M$ at $x$ and
  $K_\eps$ be the  translation by $\eps$ of $K$ along the common 
  normal of the  hyperplanes such that the medial hyperplane of
  $K_\eps$ touches the ball $E_\eps$ on the opposite of $x$. Then,
  for $\eps$ small enough, $\nr{p_K -  p_{K'}}_{\L^1(E_\eps)}
  \simeq R \times \H^n(E_\eps)$, which clearly exceeds the assumed
  bound for a small enough $\eps$.

\item According to this theorem, the map $K \mapsto p_K \in \L^1(E)$ is
  locally $1/2$-Hölder. The following example shows that this result
  cannot be improved even  around a very simple compact set.  

\begin{figure}[h]
\centering \includegraphics[height=2.5cm]{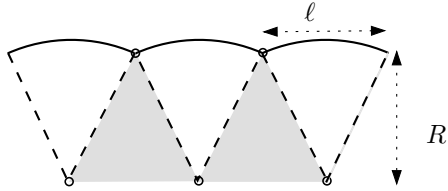}
\caption{A sequence of «knife blades» converging to a segment.}
\label{fig:knife}

\end{figure}

Let $S$ and $S'$ be two opposite sides of a
rectangle $E$, \ie two segments of length $L$ and at distance
$R$. We now define a Hausdorff approximation of $S$: for any
positive integer $N$, divide $S$ in $N$ small segments $s^i$ of
common length $\ell$, and  let $C_i$ be the unique circle with
center in $S'$ which contains the two endpoints of $s^i$. We now let
$S_N$ be the union of the circle arcs of $C_i$ comprised between
the two endpoints of $s^i$.

Then it is not very hard to see that if $R_\eps = R + \eps$ is the
common radius of all the $C^i$, $R_\eps^2 = R^2 + (\ell/2)^2$, \ie
$ \d_H(S,S_N) = \sqrt{R^2 + (\ell/2)^2} - R \leq R\ell^2/8$. Then  the
$\L^1$-distance between the projections on $S$ and $S_N$ is at least
$\Omega(\ell)$ (because almost half of the points in $E$ projects on
the corners of $S_N$, see the shaded area in
fig. \ref{fig:knife}). Hence, 
$$\nr{p_S - p_{S_N}}_{\L^1(E)} = \Omega(\ell) =
\Omega(\d_H(S,S_N)^{1/2})$$
\end{enumerate}
\end{remarks}

\subsubsection*{Replacing $\L^1(E)$ with
  $\L^1(\mu)$ where $\mu$ has bounded variation}
As we have seen before, a corollary of the previous result is that if
$\mu = \restr{\H^n}{E}$, the map $K \mapsto \p_{K\#} \mu$ is locally
$1/2$-Hölder. This result can be generalized when $\mu = u \H^n$
where $u \in \Lloc(\R^n)$ has \emph{bounded variation}. We recall some
facts about the theory of functions with bounded variation, taken from
\cite{ambrosio2000fbv}. If $\Omega \subseteq \R^n$ is an open set and
$u \in \Lloc(\Omega)$, the \emph{variation} of $u$ in $\Omega$ is
$$ \var(u,\Omega) = \sup \left\{ \int_\Omega u \ddiv \phi; \phi \in
\Ck^1_c(\Omega), \nr{\phi}_\infty \leq 1\right\} $$

A function $u \in \Lloc(\Omega)$ has \emph{bounded variation} if
$V(u,\Omega) < +\infty$. The set of functions of bounded variation on
$\Omega$ is denoted by $\BV(\Omega)$. 
We also mention that if $u$ is Lipschitz on $\Omega$, then
$\var(u,\Omega) = \nr{\nabla u}_{\L^1(\Omega)}$. Finally, we let
$\var(u)$ be the total variation of $u$ in $\R^n$.


 
\begin{theorem}
\label{th:stab:lip}
Let $\mu \in \M(\R^n)$ be a measure with density $u \in
\BV(\R^n)$ with respect to the Lebesgue measure, and $K$ be a
compact subset of $\R^n$. We suppose that $\supp(u) \subseteq
K^{R}$. Then, if $\d_H(K,K')$ is small enough,
$$ \dbL(\p_{K\#} \mu, \p_{K'\#} \mu) \leq C_2(n)
\left(\nr{u}_{\L^1(K^R)} + \diam(K) \var\left(u\right) \right)
\sqrt{R}\times \d_H(K,K')^{1/2} $$
\end{theorem}

\begin{proof}
We begin with the additional assumption that $u$ has class $\Ck^\infty$.
The function $u$ can be written as an integral over $t \in \R$ of the 
characteristic functions of its superlevel sets $E_t = \{ u > t\}$, \ie
 $u(x) = \int_0^\infty \chi_{E_t}(x) \d t$. Fubini's theorem then
 ensures that for any Lipschitz function $f$ defined on $\R^n$ with
 $\nr{f}_{\SLip} \leq 1$, 

$$
\begin{aligned}
\p_{K'\#} \mu (f) &=  \int_{\R^n} f\circ \p_{K'}(x) u(x) \d x \\
       &= \int_\R \int_{\R^n} f\circ \p_{K'}(x) \chi_{\{u \geq t\}}(x)
       \d x \d t
\end{aligned}$$

By Sard's theorem, for almost any $t$, $\partial E_t = u^{-1}(t)$ is a
$(n-1)$-rectifiable subset of $\R^n$. Thus, for those $t$ the previous
corollary implies, for $\eps = \d_H(K,K') \leq \eps_0 = \min(R,
\diam(K), \diam(K)^2/R_K)$,
$$\begin{aligned}
\int_{E_t}\abs{  f \circ \p_K(x) - f \circ \p_{K'}(x) } \d x
 &\leq \nr{\p_K - \p_{K'}}_{\L^1(E_t)}\\
&\leq C_2(n)[ \H^n(E_t) + \diam(K)
\H^{n-1}(\partial E_t)] \sqrt{R \eps}\end{aligned}$$
Putting this inequality into the last equality gives
$$ \abs{p_{K\#} \mu(f) - p_{K'\#} \mu(f)} 
\leq C_2(n) \left(\int_\R \H^n(E_t) + \diam(K) \H^{n-1}(\partial
  E_t) \d t\right) \sqrt{R\eps} $$
Using Fubini's theorem again and the coarea formula one finally gets
that   
$$\abs{p_{K\#} \mu(f) - p_{K'\#} \mu(f)} \leq C_2(n)
\left(\nr{u}_{\L^1(K^R)} + \diam(K) \var(u)
\right) \sqrt{R\eps}.$$

This proves the theorem in the case of Lipschitz functions. To
conclude the proof in the general case, one has to approximate the
bounded variation function $u$ by a sequence of $\Ck^\infty$ functions
$(u_n)$ such that both $\nr{u - u_n}_{\L^1(K^R)}$ and $\abs{\var(u) -
  \var(u_n)}$ converge to zero, which is possible by theorem 3.9
in \cite{ambrosio2000fbv}.
\end{proof}

\begin{remark}
Taking $u = \chi_E$ where $E$ is a suitable open set shows that
theorem \ref{th:stab:E} can also be recovered from \ref{th:stab:lip}.
\end{remark}

\section{Stability of boundary and curvature measures}

We combine the results of corollaries \ref{coro:triangle},
\ref{coro:szsym} and \ref{coro:diffp} to get 

\begin{theorem}
\label{th:stab}
If $K$ and $K'$ are two compact sets with $\eps = \d_H(K,K')$ smaller
than $\min(\diam K,r,r^2/\diam K)$, then
$$ \dbL(\mu_{K,r}, \mu_{K',r}) \leq 
C_3(n)\NN(K,r-\eps) r^{n} [r + \diam(K)] \sqrt{\frac{\eps}{r}}$$

In particular, if for a given bounded Lipschitz function $f$ on
$\R^n$, one defines $\phi_{K,f}(r) = \mu_{K,r}(f)$, the map $ K
\mapsto \phi_{K,f} \in \Ck^0([r_\mathrm{min},r_\mathrm{max}])$  with 
$0 < r_\mathrm{min} < r_\mathrm{max}$ is
locally $1/2$-Hölder. 
\end{theorem}

In what follows we suppose that $(r_i)$ is a sequence of $n$ distinct
numbers $0 < r_0 < ... < r_{n}$. For any compact set $K$ and $f \in
\Ck^0(\R^n)$, we let $\left[\Phi^{(r)}_{K,i}(f)\right]_i$ be the solutions of
the linear system 
$$\forall i \hbox{ s.t } 0\leq i \leq n,~ \sum_{j=0}^n \omega_{n-j}
\Phi^{(r)}_{K,j}(f) r_i^{n-j} = \mu_{K,r_i}(f)$$
Since the system is linear in $(\mu_{K,r_i}(f))$ and these values
depends continuously on $f$, the map $f \mapsto \Phi_{K,i}^{(r)}(f)$
is also linear and continuous, \ie $\Phi^{(r)}_{K,i}$ is a signed measure on
$\R^n$. It is also to be noticed that if $K$ has positive reach with
$\reach(K) > r_n$, the $\Phi^{(r)}_{K,i}$ coincide with the usual
curvature measures of $K$. In that case, the following result gives a
way to approximate the (usual) curvature measures of $K$ from a
Hausdorff-approximation of it even if its reach is arbitrary
small.

\begin{corollary}
There exist a constant $C$ depending on $K$ and $(r)$ such that
for any compact subset $K'$ of $\R^n$ close enough to $K$,
$$ \forall i,~ \dbL\left(\Phi_{K',i}^{(r)}, \Phi_{K,i}^{(r)}\right) \leq C
\d_H(K,K')^{1/2} $$
\end{corollary}

\bibliography{RR-dim}
\bibliographystyle{alpha}

\end{document}